\documentclass[aps,prd,reprint,groupedaddress,superscriptaddress,nofootinbib]{revtex4-1}

\usepackage{graphicx}
\usepackage{amsmath,amssymb}             % AMS Math
\usepackage{hyperref}

\newcommand{\arXivred}[1]{\href{http://arxiv.org/abs/#1}{arXiv:#1}}
\newcommand{\arXiv}[2]{\href{http://arxiv.org/abs/#1}{arXiv:#1} [#2]}

\begin{document}

%\preprint{}

\title{Bayesian inference of dark matter voids in galaxy surveys}


\author{Florent Leclercq}
\email{florent.leclercq@polytechnique.org}
%\homepage[]{test}
%\thanks{}
\affiliation{Institut d'Astrophysique de Paris (IAP), UMR 7095, CNRS - UPMC Universit\'e Paris 6, 98bis boulevard Arago, F-75014 Paris, France}
\affiliation{Institut Lagrange de Paris (ILP), Sorbonne Universit\'es, 98bis boulevard Arago, F-75014 Paris, France}
\affiliation{\'Ecole polytechnique ParisTech, Route de Saclay, F-91128 Palaiseau, France}


\date{\today}

\begin{abstract}
\noindent We apply the \textsc{borg} algorithm to the Sloan Digital Sky Survey Data Release 7 main sample galaxies. The method results in the physical inference of the initial density field at a scale factor $a~=~10^{-3}$, evolving gravitationally to the observed density field at a scale factor $a~=~1$, and provides an accurate quantification of corresponding uncertainties.
Building upon these results, we generate a set of constrained realizations of the present large-scale dark matter distribution.
As a physical illustration, we apply a void identification algorithm to them. In this fashion, we access voids defined by the inferred dark matter field, not by galaxies, greatly alleviating the issues due to the sparsity and bias of tracers. In addition, the use of full-scale physical density fields yields a drastic reduction of statistical uncertainty in void catalogs. These new catalogs are enhanced data sets for cross-correlation with other cosmological probes.
\end{abstract}

%\pacs{67886-A656}
%\keywords{}

\maketitle


% -----------------------------------------------------------------------------
% -----------------------------------------------------------------------------

\section{Bayesian physical inference of the initial conditions}

We apply~\cite{JLW2014} the full-scale Bayesian inference code \textsc{borg} (Bayesian Origin Reconstruction from Galaxies,~\cite{JascheWandelt2013}) to the galaxies of the \texttt{Sample dr72} of the New York University Value Added Catalogue (NYU-VAGC~\footnote{\href{http://sdss.physics.nyu.edu/vagc/}{http://sdss.physics.nyu.edu/vagc/}}), based on the final data release (DR7) of the Sloan Digital Sky Survey (SDSS). The physical model for gravitational dynamics is second-order Lagrangian perturbation theory (2LPT), linking initial density fields ($a~=~10^{-3}$) to the presently observed large-scale structure, in the linear and mildly non-linear regime. The algorithm explores numerically the posterior distribution by sampling the joint distribution of all parameters involved, via efficient Hamiltonian Markov Chain Monte Carlo (HMC) dynamics.

\begin{figure*}
\begin{center}
\includegraphics[width=\textwidth]{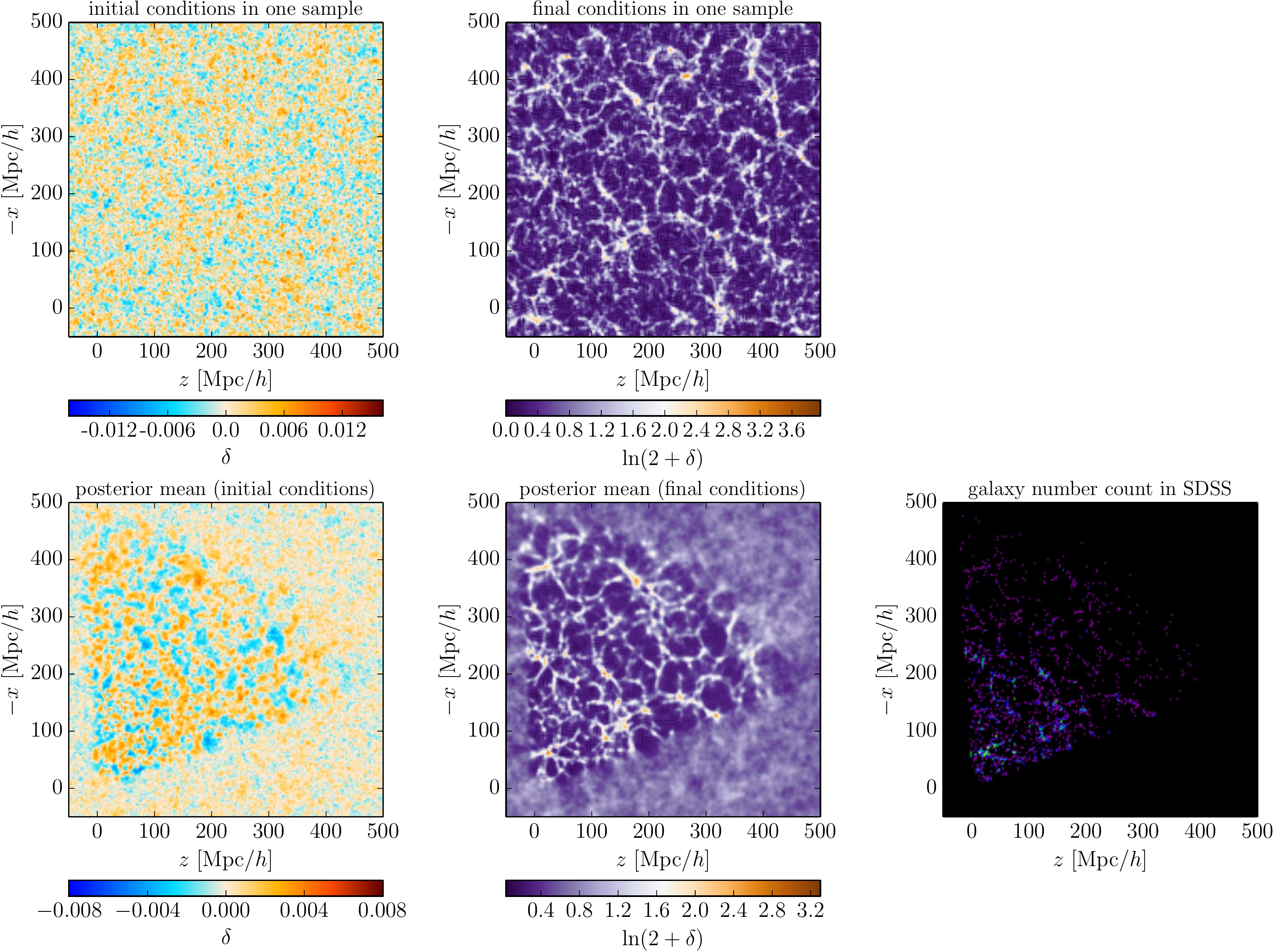} 
\end{center}
\caption{Bayesian large-scale structure inference with \textsc{borg}. Upper panel: slices through one sample of the posterior for the initial (left) and final (center) density fields. Lower panel: posterior mean in the initial (left) and final (center) conditions, compared to the input data (right).}
\label{fig:slices}
\end{figure*}

Each sample (Fig. \ref{fig:slices}, upper panel) is a ``possible version of the truth'' in the form of a full physical realization of dark matter particles, tracing both the density and the velocity fields. The variation between samples (Fig. \ref{fig:slices}, lower panel) quantifies joint and correlated uncertainties (survey geometry, selection effects, biases, noise) inherent to any cosmological observation. 

\section{Data-contrained realizations of the Universe}

\begin{figure*}
\begin{center}
\includegraphics[width=\textwidth]{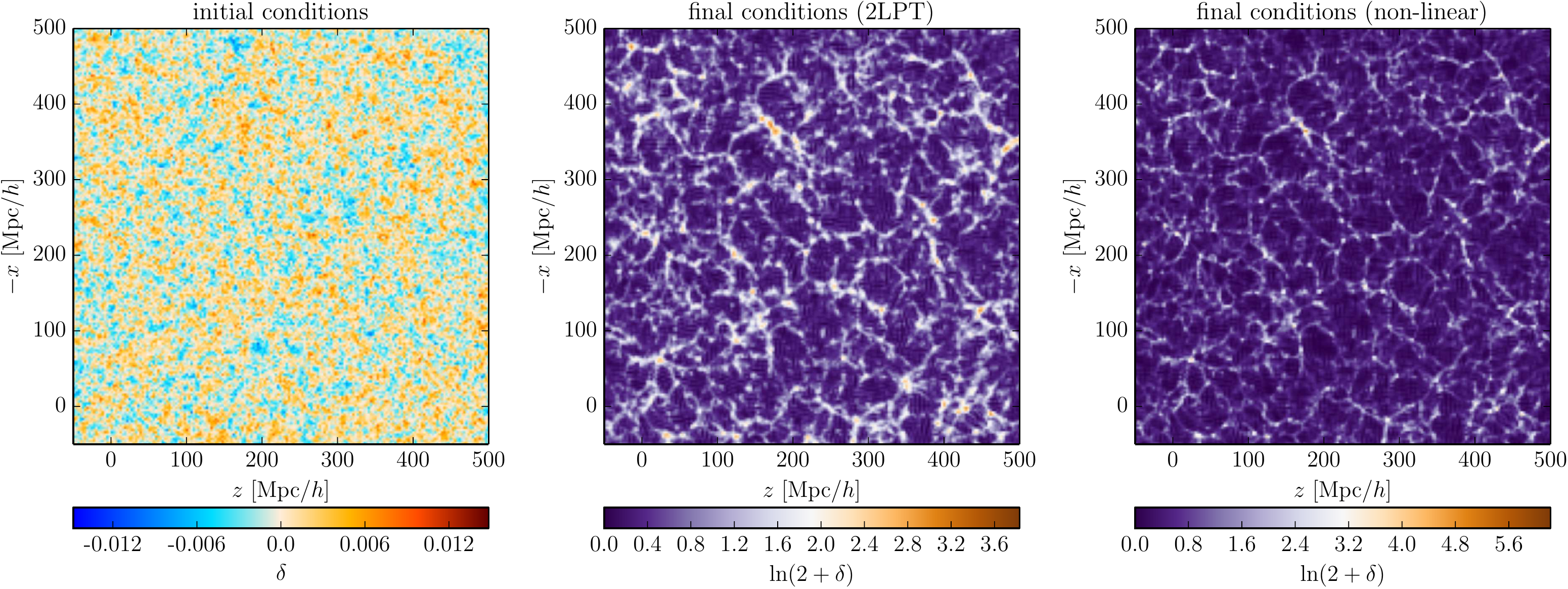} 
\end{center}
\caption{Non-linear filtering of \textsc{borg} results. Slices through one sample of initial (left panel) and final density fields (middle panel) inferred by \textsc{borg}. The final density field (middle panel) is a prediction of the 2LPT model used by \textsc{borg}. On the right panel, a slice through the data-constrained realization obtained with the same sample via non-linear filtering (fully non-linear gravitational structure formation starting from the same initial conditions) is shown.}
\label{fig:slices_filter}
\end{figure*}

We generate~\cite{LJSHW2014} a set of data-constrained realizations of the present large-scale structure (Fig. \ref{fig:slices_filter}): some samples of inferred primordial conditions are evolved with 2LPT to $z=69$, then with a fully non-linear cosmological simulation (using \textsc{gadget-2},~\cite{Springel2005}) from $z=69$ to $z=0$.

A dynamic, non-linear physical model naturally introduces some correlations between the constrained and unconstrained parts, which yields reliable extrapolations for certain aspects of the model that have not yet been constrained by the data (e.g. near the survey boundaries or at high redshift). 

\section{Dark matter voids in the SDSS galaxy survey}

We apply~\cite{LJSHW2014} \textsc{vide} (the Void IDentification and Examination toolkit,~ \cite{VIDE}~\footnote{\href{http://www.cosmicvoids.net/}{http://www.cosmicvoids.net/}}) to the constrained parts of these realizations. The void finder is a modified version of \textsc{zobov}~\cite{Neyrinck2008} that uses Voronoi tessellations of the tracer particles to estimate the density field and a watershed algorithm to group Voronoi cells into voids.

\begin{figure*}
\begin{center}
\includegraphics[width=0.48\textwidth]{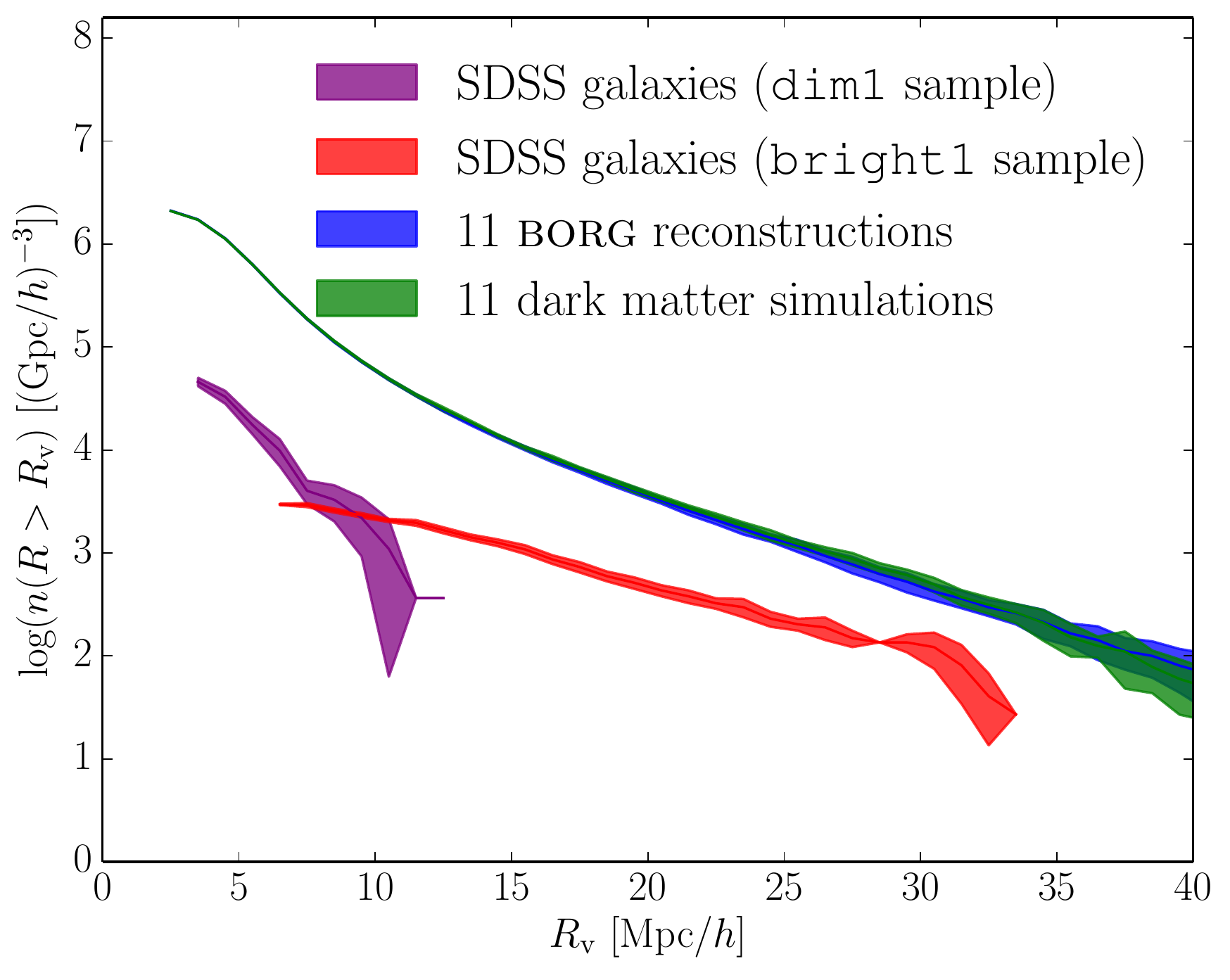} \quad \includegraphics[width=0.48\textwidth]{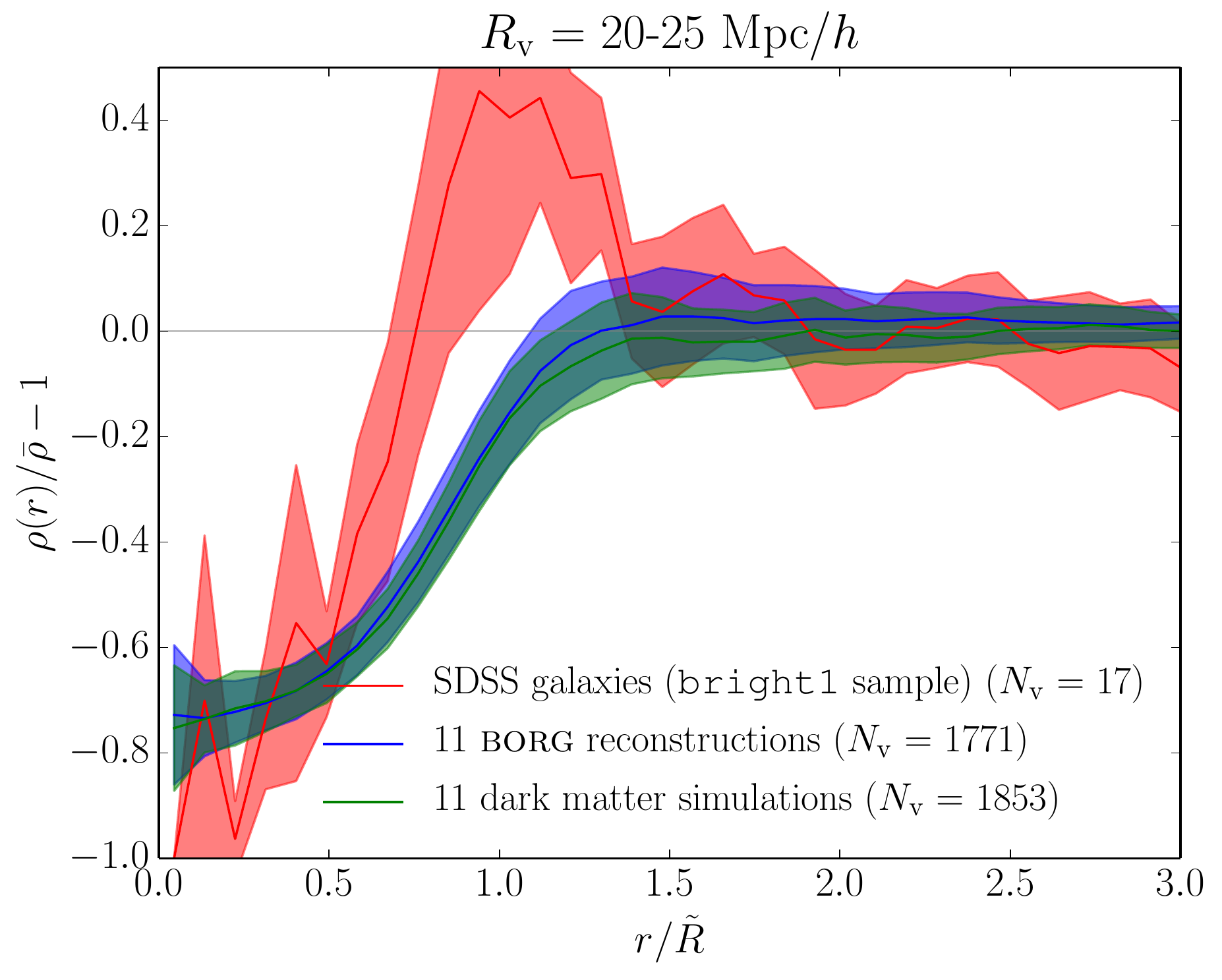}
\end{center}
\caption{Left panel: void number count in 11 \textsc{borg} reconstructions (blue), compared to the voids found in $N$-body simulations prepared with the same setup (green) and to the voids directly traced by the SDSS galaxies (red and purple). Right panel: density profile for stacked voids of radius between 20 and 25 Mpc/$h$ in the same void catalogs.}
\label{fig:voids}
\end{figure*}

We find physical cosmic voids in the field traced by the dark matter particles, probing a level deeper in the mass distribution hierarchy than galaxies, and greatly alleviating the bias problem for cosmological interpretation of final results. Due to the high density of tracers, we find about an order of magnitude more voids at all scales than the voids directly traced by the SDSS galaxies (Fig. \ref{fig:voids}, left panel), which sample the underlying mass distribution only sparsely~\cite{Sutter2014}. Our inference framework therefore yields a drastic reduction of statistical uncertainty in voids catalogs. For usual voids statistics such as radial density profiles of stacked voids (observed in simulations to be of universal character, e.g.~\cite{Hamaus2014}), the results we obtain are consistent with $N$-body simulations prepared with the same setup (Fig. \ref{fig:voids}, right panel).

\section*{Acknowledgments}

I thank Nico Hamaus, Jens Jasche, Paul Sutter and Benjamin Wandelt for fruitful collaborations and useful discussions. I acknowledge funding from an AMX grant (\'Ecole polytechnique ParisTech) and Benjamin Wandelt's senior Excellence Chair by the Agence Nationale de la Recherche (ANR-10-CEXC-004-01). This work made in the ILP LABEX (ANR-10-LABX-63) was supported by French state funds managed by the ANR within the Investissements d'Avenir programme (ANR-11-IDEX-0004-02).

\end{document}